\documentclass[10pt,conference]{IEEEtran}
\IEEEoverridecommandlockouts
\pdfoutput=1
\usepackage{cite}
\usepackage{amsmath,amssymb,amsfonts}
\usepackage{algorithmic}
\usepackage{algorithm}
\usepackage{graphicx}
\usepackage{textcomp}
\usepackage{xcolor}
\usepackage{units}
\usepackage{balance}

\usepackage{tabu,longtable}
\usepackage{diagbox}
\usepackage{threeparttable}
\usepackage{color}

\DeclareMathOperator*{\argmin}{arg\,min}

\usepackage{tikz} % for tikz
\usepackage[utf8]{inputenc}
\usepackage{pgfplots} 
\usepackage{pgfgantt}
\usepackage{pdflscape}
\usepackage{bm}
\usepackage{ulem}

\usepackage{acro}
\DeclareAcronym{5g}{
short=5G,
long= fifth generation,
}
\DeclareAcronym{6g}{
short=6G,
long= sixth generation,
}
\DeclareAcronym{3d}{
short=3D,
long= three-dimensional,
}
\DeclareAcronym{2d}{
short=2D,
long= two-dimensional,
}
\DeclareAcronym{aod}{
short=AOD,
long= angle-of-departure,
}
\DeclareAcronym{aosa}{
short=AOSA,
long= array-of-subarray,
}
\DeclareAcronym{adod}{
short=ADOD,
long= angle-difference-of-departure,
}
\DeclareAcronym{aoa}{
short=AOA,
long= angle-of-arrival,
}
\DeclareAcronym{adc}{
short=ADC,
long= analog to digital converter,
}
\DeclareAcronym{aeb}{
short=AEB,
long= angle error bound,
}
\DeclareAcronym{av}{
short=AV,
long= autonomous vehicle,
}
\DeclareAcronym{bs}{
short=BS,
long= base station,
}

\DeclareAcronym{csi}{
short=CSI,
long= channel state information,
}
\DeclareAcronym{cfo}{
short=CFO,
long= carrier frequency offset,
}
\DeclareAcronym{ceb}{
short=CEB,
long= clock error bound,
}
\DeclareAcronym{coa}{
short=COA,
long= curvature-of-arrival,
}
\DeclareAcronym{crb}{
short=CRB,
long= Cram\'er-Rao bound,
}
\DeclareAcronym{ccrb}{
short=CCRB,
long= constrained Cram\'er-Rao bound,
}
\DeclareAcronym{cmos}{
short=CMOS,
long= complementary metal-oxide-semiconductor,
}
\DeclareAcronym{crlb}{
short=CRLB,
long= Cram\'er-Rao lower bound,
}
\DeclareAcronym{cdf}{
short=CDF,
long= cumulative distribution function,
}
\DeclareAcronym{cp}{
short=CP,
long= cyclic prefix,
}
\DeclareAcronym{dac}{
short=DAC,
long= digital to analog converter,
}
\DeclareAcronym{dfl}{
short=DFL,
long= device-free localization,
}
\DeclareAcronym{dmimo}{
short=D-MIMO,
long= distributed MIMO,
}
\DeclareAcronym{dlprs}{
short=DL-PRS,
long= downlink positioning reference signal,
}
\DeclareAcronym{d2d}{
short=D2D,
long= device-to-device,
}
\DeclareAcronym{dftsofdm}{
short=DFT-s-OFDM,
long= discrete-Fourier-transform spread OFDM,
}
\DeclareAcronym{dl}{
short=DL,
long= deep learning,
}
\DeclareAcronym{gps}{
short=GPS,
long= global positioning system,
}
\DeclareAcronym{gnss}{
short=GNSS,
long= global navigation satellite system,
}
\DeclareAcronym{fim}{
short=FIM,
long = Fisher information matrix,
}
\DeclareAcronym{hwi}{
short=HWI,
long= hardware impairment,
}
\DeclareAcronym{hemt}{
short=HEMT,
long= high electron mobility transistor,
}
\DeclareAcronym{hbt}{
short=HBT,
long= heterojunction bipolar transistors,
}
\DeclareAcronym{iot}{
short=IoT,
long= internet of things,
}
\DeclareAcronym{imu}{
short=IMU,
long= inertial measurement unit,
}
\DeclareAcronym{isac}{
short=ISAC,
long= integrated sensing and communication,
}
\DeclareAcronym{iqi}{
short=IQI,
long= in-phase and quadrature imbalance,
}
\DeclareAcronym{ip}{
short=IP,
long= incidence point,
}
\DeclareAcronym{ia}{
short=IA,
long= initial access,
}
\DeclareAcronym{kpi}{
short=KPI,
long= key performance indicator,
}
\DeclareAcronym{kf}{
short=KF,
long= Kalman filter,
}
\DeclareAcronym{ekf}{
short=EKF,
long= extended Kalman filter,
}
\DeclareAcronym{ukf}{
short=UKF,
long= unscented Kalman filter,
}
\DeclareAcronym{ckf}{
short=CKF,
long= cubature Kalman filter,
}
\DeclareAcronym{pf}{
short=PF,
long= particle filter,
}
\DeclareAcronym{lb}{
short=LB,
long= lower bound,
}
\DeclareAcronym{ls}{
short=LS,
long= least-squares,
}
\DeclareAcronym{lo}{
short=LO,
long= local oscillator,
}
\DeclareAcronym{los}{
short=LOS,
long= line-of-sight,
}
\DeclareAcronym{mc}{
short=MC,
long= mutual coupling,
}
\DeclareAcronym{mac}{
short=MAC,
long= medium access control,
}
\DeclareAcronym{meb}{
short=MEB,
long= mapping error bound,
}
\DeclareAcronym{ml}{
short=ML,
long= machine learning,
}
\DeclareAcronym{mcrb}{
short=MCRB,
long= misspecified Cram\'er-Rao bound,
}
\DeclareAcronym{mds}{
short=MDS,
long= multidimensional scaling ,
}
\DeclareAcronym{mimo}{
short=MIMO,
long= multiple-input-multiple-output,
}
\DeclareAcronym{siso}{
short=SISO,
long= single-input-single-output,
}
\DeclareAcronym{mm}{
short=MM,
long= mismatched model,
}
\DeclareAcronym{mpc}{
short=MPC,
long= multipath components,
}
\DeclareAcronym{mmwave}{
short=mmWave,
long= millimeter wave,
}
\DeclareAcronym{mmle}{
short=MMLE,
long= mismatched maximum likelihood estimation,
}
\DeclareAcronym{mems}{
short=MEMS,
long= micro-electro-mechanical system,
}
\DeclareAcronym{mle}{
short=MLE,
long= maximum likelihood estimation,
}
\DeclareAcronym{nlos}{
short=NLOS,
long= none-line-of-sight,
}
\DeclareAcronym{ofdm}{
short=OFDM,
long= orthogonal frequency-division multiplexing,
}
\DeclareAcronym{oeb}{
short=OEB,
long= orientation error bound,
}
\DeclareAcronym{otfs}{
short=OTFS,
long= orthogonal time-frequency space,
}
\DeclareAcronym{pdf}{
short=PDF,
long= probability density function,
}
\DeclareAcronym{papr}{
short=PAPR,
long= peak-to-average-power ratio,
}
\DeclareAcronym{pan}{
short=PAN,
long= power amplifier nonlinearity,
}
\DeclareAcronym{pa}{
short=PA,
long= power amplifier,
}
\DeclareAcronym{ps}{
short=PS,
long= phase shifter,
}
\DeclareAcronym{pn}{
short=PN,
long= phase noise,
}
\DeclareAcronym{poa}{
short=POA,
long= phase-of-arrival,
}
\DeclareAcronym{pwm}{
short=PWM,
long= planar wave model,
}
\DeclareAcronym{pdoa}{
short=PDOA,
long= phase-difference-of-arrival,
}
\DeclareAcronym{prs}{
short=PRS,
long= positioning reference signals,
}
\DeclareAcronym{peb}{
short=PEB,
long= position error bound,
}
\DeclareAcronym{rnn}{
short=RNN,
long= recurrent neural network,
}
\DeclareAcronym{rl}{
short=RL,
long= reinforcement learning,
}
\DeclareAcronym{rfc}{
short=RFC,
long= radio-frequency chain,
}
\DeclareAcronym{rf}{
short=RF,
long= radio frequency,
}
\DeclareAcronym{rfid}{
short=RFID,
long= radio frequency identification,
}
\DeclareAcronym{ris}{
short=RIS,
long= reconfigurable intelligent surface,
}
\DeclareAcronym{rss}{
short=RSS,
long= received signal strength,
}
\DeclareAcronym{rtt}{
short=RTT,
long= round-trip time,
}
\DeclareAcronym{sm}{
short=SM,
long= standard model,
}
\DeclareAcronym{sige}{
short=SiGe,
long= silicon-germanium,
}
\DeclareAcronym{spp}{
short=SPP,
long= surface plasmon polariton,
}
\DeclareAcronym{sa}{
short=SA,
long= subarray,
}
\DeclareAcronym{sota}{
short=SOTA,
long= state-of-the-art,
}
\DeclareAcronym{swm}{
short=SWM,
long= spherical wave model,
}
\DeclareAcronym{slam}{
short=SLAM,
long= simultaneous localization and mapping,
}
\DeclareAcronym{tm}{
short=TM,
long= true model,
}
\DeclareAcronym{toa}{
short=TOA,
long= time-of-arrival,
}
\DeclareAcronym{tof}{
short=TOF,
long= time-of-flight,
}
\DeclareAcronym{tdoa}{
short=TDOA,
long= time-difference-of-arrival,
}
\DeclareAcronym{thz}{
short=THz,
long= Terahertz,
}
\DeclareAcronym{ue}{
short=UE,
long= user equipment,
}
\DeclareAcronym{ummimo}{
short=UM-MIMO,
long= ultra-massive multi-input-multi-output,
}
\DeclareAcronym{vlp}{
short=VLP,
long= visible light positioning,
}
\DeclareAcronym{veb}{
short=VEB,
long= velocity error bound,
}
\DeclareAcronym{vlc}{
short=VLC,
long= visible light communication,
}
\DeclareAcronym{ula}{
short=ULA,
long= uniform linear array,
}
\DeclareAcronym{uav}{
short=UAV,
long= unmanned aerial vehicle,
}
\DeclareAcronym{upa}{
short=UPA,
long= uniform planar array,
}
\DeclareAcronym{wlan}{
short=WLAN,
long= wireless local area network,
}
\DeclareAcronym{rtk}{
short = RTK,
long = real-time kinematic,
}
\DeclareAcronym{rmse}{
short=RMSE,
long= root mean square error,
}

\newcommand{\TT}{\mathsf{T}}

\newcommand{\bv}{{\bf b}}

\newcommand{\hv}{{\bf h}}

\newcommand{\kv}{{\bf k}}

\newcommand{\pv}{{\bf p}}

\newcommand{\tv}{{\bf t}}
\newcommand{\uv}{{\bf u}}

\newcommand{\xv}{{\bf x}}
\newcommand{\yv}{{\bf y}}

\newcommand{\zerov}{{\bf 0}}

\newcommand{\Am}{{\bf A}}
\newcommand{\Bm}{{\bf B}}
\newcommand{\Cm}{{\bf C}}

\newcommand{\Hm}{{\bf H}}
\newcommand{\Id}{{\bf I}}

\newcommand{\Qm}{{\bf Q}}
\newcommand{\Rm}{{\bf R}}

\newcommand{\Wm}{{\bf W}}

\newcommand{\Zerom}{{\bf 0}}
\newcommand{\Bt}{{\rm B}}

\newcommand{\phiv}{\hbox{\boldmath$\phi$}}

\newcommand{\thetav}{\hbox{$\boldsymbol\theta$}}
\newcommand{\tauv}{\hbox{\boldmath$\tau$}}

\begin{document}

\bstctlcite{IEEEexample:BSTcontrol}

\title{
5G-Aided RTK Positioning in GNSS-Deprived Environments\\
\author{
Pinjun Zheng, 
Xing Liu,
Tarig Ballal, 
Tareq Y. Al-Naffouri\\
\textit{Computer, Electrical and Mathematical Science \& Engineering}\\
\textit{King Abdullah University of Science and Technology (KAUST), Thuwal, KSA}\\
Email: \{pinjun.zheng; xing.liu; tarig.ahmed; tareq.alnaffouri\}@kaust.edu.sa
}

}

\maketitle

\begin{abstract}
This paper considers the localization problem in a 5G-aided global navigation satellite system (GNSS) based on real-time kinematic (RTK) technique.
Specifically, the user's position is estimated based on the hybrid measurements, including GNSS pseudo-ranges, GNSS carrier phases, 5G angle-of-departures, and 5G channel delays.
The underlying estimation problem is solved by steps that comprise obtaining the float solution, ambiguity resolution, and resolving the fixed solution.
The analysis results show that the involvement of 5G observations can enable localization under satellite-deprived environments, inclusive of extreme cases with only 2 or 3 visible satellites.
Moreover, extensive simulation results reveal that with the help of 5G observations, the proposed algorithm can significantly reduce the estimation error of the user's position and increase the success rate of carrier-phase ambiguity resolution.
\end{abstract}

\begin{IEEEkeywords}
GNSS, 5G/6G, localization, RTK, ambiguity resolution.
\end{IEEEkeywords}

\section{Introduction}

Accurate localization has become an essential requirement for a broad variety of applications, such as intelligent transportation, precision agriculture, surveying and mapping, and smart cities~\cite{7904784,9163228,8347072}. 
The development of advanced unmanned systems in recent years motivates a further increase in the demand for high positioning accuracy and reliability~\cite{yuan2021survey,chen2021trends}. 
Although plenty of navigation techniques have been developed for outdoor positioning, \ac{gnss}-based positioning is the most prevalent thanks to its advantages of high accuracy, global coverage, low cost, and all-weather capability. 

\Ac{rtk} is a widely-used \ac{gnss}-based positioning technique. In \ac{rtk} positioning, a \ac{gnss} \ac{bs} installed at a fixed position and the user's \ac{gnss} receiver collect \ac{gnss} observations simultaneously. The \ac{bs} transmits its observation (the pseudo-range and carrier-phase) together with its accurate position to the user via a suitable communication link~\cite{jeffrey2010introduction}. 
The involvement of \ac{gnss} carrier-phase observations, differential correction, and ambiguity resolution enables RTK positioning to provide centimeter-level accuracy in open-sky scenarios~\cite{parkins2011increasing, hofmann2007gnss}. 
However, the performance of \ac{rtk} in deep urban environments is not up to par with the high-accuracy requirements for many dynamic systems. In such environments, buildings can block, weaken, reflect, and diffract the \ac{gnss} signals, which may result in an insufficient number of visible satellites and observations with severe multipath effects~\cite{liu2019gnss}. 
Providing reliable navigation in these situations is a daunting task that is yet to be accomplished. 

To address the limitations of \ac{rtk} positioning, fusing various sensor types (including inertial navigation systems, optical sensors, Lidar, etc.) with \ac{gnss} has been explored ~\cite{li2018high,schutz2020precise}. 
Recently, with the emergence of the \ac{5g} wireless systems which are expected to provide high-precision localization services, many promising results about \ac{5g} localization have been reported in the literature~\cite{Koivisto2017High,Koivisto2017Joint,Shahmansoori2018Position,De2021Convergent}. 
To take advantage of these emerging technologies, \ac{gnss} augmentation with \ac{5g} has been considered. 
Examples of such works include hybrid \ac{gnss}-\ac{5g} positioning  based on device-to-device measurements~\cite{Yin2018GNSS}, neural network fingerprinting and \ac{gnss} data fusion~\cite{Klus2021Neural}, multi-rate \ac{5g} and \ac{gnss} data fusion~\cite{Bai2022GNSS}, and a few more~\cite{Destino2018Performance,Minetto2022DGNSS}.

This paper utilizes \ac{5g} observations to aid \ac{rtk} positioning to overcome its shortcomings in GNSS-deprived environments.
Our main contribution is proposing a method to leverage \ac{5g} observations in GNSS carrier-phase ambiguity resolution, especially in harsh environments with poor satellite visibility.
We formulate the localization problem by jointly using \ac{gnss} and \ac{5g} observations. 
The problem is solved through three steps, namely, computing a float solution, ambiguity resolution, and computing the fixed solution.
In addition, we perform a localization availability analysis which demonstrates that the introduction of the \ac{5g} observations can enable localization in extreme scenarios with only 2 or 3 visible satellites.

\section{System Model}
\begin{figure}[t]
    \centering
    \includegraphics[width=3.45in]{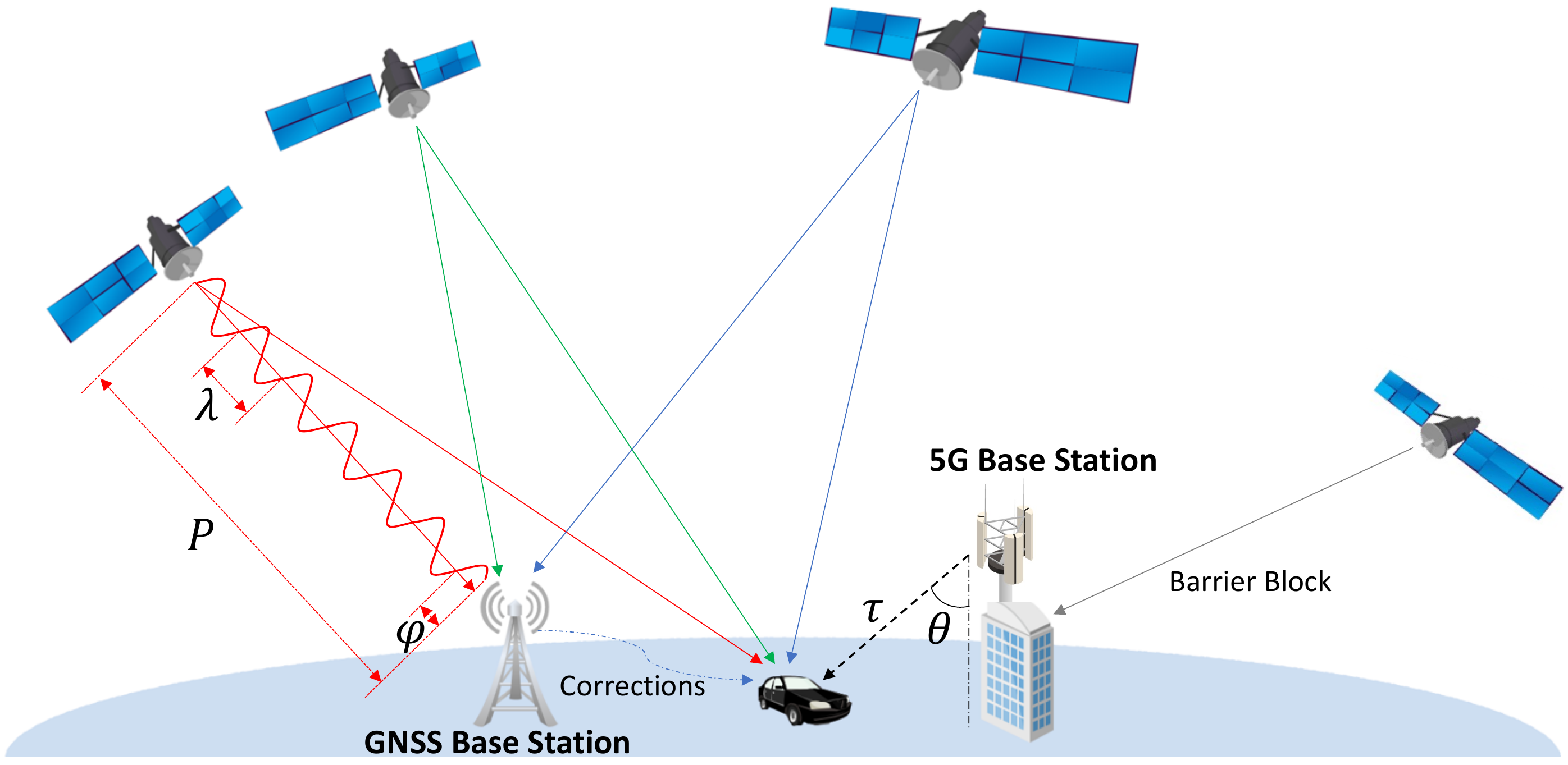}
    \caption{ 
        Illustration of a \ac{5g}-aided \ac{rtk} positioning system with three satellites and one \ac{5g} \ac{bs}. 
      }
    \label{fig_system}
\end{figure}

We consider a \ac{5g}-aided \ac{rtk} positioning system with $N$ visible satellites, a user equipment with an unknown position, a \ac{gnss} \ac{bs} with a known position, and $L$ \ac{5g} \acp{bs} with known positions and orientations. A \ac{5g} radio link is established between the \ac{gnss} \ac{bs} and the user. 
The user can receive both \ac{gnss} and \ac{5g} signals. 
An example of the considered system with $N=3$ and $L=1$ is illustrated in Fig.~\ref{fig_system}.
For clarity, we use superscripts $(\cdot)^n$ to denote the variables related to the $n$-th satellite, 
subscripts $(\cdot)_u$, $(\cdot)_b$ and $(\cdot)_\Bt$ represent the variables related to the user, the \ac{gnss} \ac{bs} and the \ac{5g} \ac{bs}, respectively.
We denote the user position as $\pv_u\in\mathbb{R}^3$ and the position of the \ac{gnss} \ac{bs} as $\pv_b\in\mathbb{R}^3$.
The position and orientation of the $\ell$-th \ac{5g} \ac{bs} are denoted as $\pv_{\Bt,\ell}\in\mathbb{R}^3$ and $\Rm_{\Bt,\ell}\in\text{SO(3)}$, respectively.
Here $\text{SO(3)}$ denotes the special orthogonal group of \ac{3d} rotation matrices
$
  \text{SO}(3) = \{\mathbf{R}|\mathbf{R}^\mathsf{T}\mathbf{R}=\mathbf{I}_3,\det(\mathbf{R}) = 1\}.
$

\subsection{RTK Model}
The main \ac{gnss} observations from the $n$-th satellite received by the user are the pseudo-range $P_u^n$ and carrier-phase $\phi_u^n$. These observations are often modeled as
\begin{align}
{P}_u^n &= \rho _u^n + I_u^n + T_u^n + c\left( {\delta {t_u} - \delta {t^n}} \right) + \varepsilon _u^n,\\
{\varphi}_u^n &= \rho _u^n + \lambda {K_u^n} - I_u^n + T_u^n + c\left( {\delta {t_u} - \delta {t^n}} \right) + \varsigma _u^n,
\end{align}
where $\rho$ denotes the geometrical range between the GNSS receiver and the satellite, $\lambda$ represents the wavelength of the \ac{gnss} carrier, $K$ represents the carrier-phase ambiguity, $I$ is the ionospheric delay, $T$ is the tropospheric delay, $\delta {t}$ is the receiver or satellite clock bias, and $\varepsilon$ and $\varsigma$ lump errors from other sources together with additive noise.

\ac{rtk} positioning takes advantage of differencing operations to virtually cancel common errors between the GNSS receivers and satellites. First, the user's and the GNSS BS's measurements collected simultaneously are differenced to eliminate the satellite clock bias and atmospheric delays, resulting in the \textit{single-difference} (SD) model \cite{teunissen2017springer}: 
\begin{equation}
\begin{aligned}
P_{ub}^n &= P_u^n - P_b^n =\rho _{ub}^n + c\delta {t_{ub}} + \varepsilon _{ub}^n,\\
\varphi_{ub}^n &= \varphi_u^n -\varphi _b^n = \rho _{ub}^n + \lambda {{K}_{ub}^n} + c\delta {t_{ub}} + \varsigma _{ub}^n. \notag
\end{aligned}
\end{equation}
Defining the unit direction vector of the satellite-user \ac{los} as $\hv_u^n$, we can obtain the following relationship based on the far-field assumption:
\begin{equation}
    \rho_{ub}^n = \rho_{u}^n - \rho_{b}^n = ({\bf{h}}_{u}^n)^\TT(\pv_u-\pv_b).
\end{equation}
Subsequently, the SD observations are again differenced over pairs of satellites to remove the receiver clock bias, which yields the \textit{double-difference} (DD) model:
\begin{equation}\notag
    \begin{aligned}
        P_{ub}^{nm} &= P_{ub}^{n} - P_{ub}^{m} =\left({\bf{h}}_{u}^n - {\bf{h}}_{u}^m\right)^\TT({\bf{p}}_{u}-{\bf{p}}_{b}) +  \varepsilon _{ub}^{nm},
    \end{aligned}
\end{equation}
\begin{equation}\notag
    \begin{aligned}
        \varphi _{ub}^{nm} &=   \varphi _{ub}^{n} -  \varphi _{ub}^{m} = \left({\bf{h}}_{u}^n - {\bf{h}}_{u}^m\right)^\TT({\bf{p}}_{u}-{\bf{p}}_{b}) + \lambda  K_{ub}^{nm} +  \varsigma _{ub}^{nm}.
     \end{aligned}
\end{equation}

Without loss of generality, we take the 1-st satellite as a reference and concatenate all the DD observations as
\begin{align}
    \pv &= [P_{ub}^{21},P_{ub}^{31},\dots,P_{ub}^{N1}]^\TT\in\mathbb{R}^{N-1},\\
    \phiv &= [\varphi_{ub}^{21},\varphi_{ub}^{31},\dots,\varphi_{ub}^{N1}]^\TT\in\mathbb{R}^{N-1}.
\end{align}
The noise-free observation model of the \ac{rtk} can be summarized as 
\begin{equation}\label{eq_RTKobs}
    \yv_1 = 
    \begin{bmatrix}
        \pv \\
        \phiv \\
    \end{bmatrix}
    =\underbrace{
    \begin{bmatrix}
        \Hm\\
        \Hm\\
    \end{bmatrix}
    }_{\Bm} \pv_u
    +
    \underbrace{
    \begin{bmatrix}
        \Zerom \\
        \lambda\Id \\
    \end{bmatrix}
    }_{\Cm}\kv
    -
    \underbrace{
    \begin{bmatrix}
        \Hm\\
        \Hm
    \end{bmatrix}
    \pv_b}_{\bv}\in\mathbb{R}^{2N-2},
\end{equation}
where 
\begin{align}\notag
    \Hm &= 
    \begin{bmatrix}
        (\hv_u^2-\hv_u^1),\dots,
        (\hv_u^{N}-\hv_u^1)
    \end{bmatrix}^\TT\in\mathbb{R}^{(N-1)\times 3}, \\
    \kv &= \begin{bmatrix}
        K_{ub}^{21},K_{ub}^{31},\dots,K_{ub}^{N1}\end{bmatrix}^\TT\in\mathbb{R}^{N-1}.\notag
\end{align}

\subsection{5G Model}

Besides the \ac{gnss} measurements, the user can receive \ac{5g} signals from $L$ \acp{bs}.
For simplicity, we assume that an \textit{efficient} channel estimator is applied and the \acp{aod} $\{\thetav_\ell\}_{\ell=1}^L$ and channel delays $\{\tau_\ell\}_{\ell=1}^L$ are available as the \ac{5g} observations~\cite{Shahmansoori2018Position}.
Note that each \ac{aod} $\thetav_\ell$ comprises an azimuth angle $\theta_\ell^{\text{az}}$ and an elevation angle $\theta_\ell^{\text{el}}$.
The \ac{5g} observation model is given by 
\begin{align}
    \theta_\ell^\text{az} &= \text{atan2}\left([\Rm_{\Bt,\ell}^\TT(\pv_u \! - \! \pv_{\Bt,\ell})]_2,[\Rm_{\Bt,\ell}^\TT(\pv_u-\pv_{\Bt,\ell})]_1\right),
    \label{eq_aodaz}\\
    \theta_\ell^\text{el} &= \text{asin}\left([\Rm_{\Bt,\ell}^\TT(\pv_u-\pv_{\Bt,\ell})]_3/\|\pv_u-\pv_{\Bt,\ell}\|_2\right),
    \label{eq_aodel}\\
    \tau_\ell &= \frac{\|\pv_{\Bt,\ell}-\pv_u\|_2}{c} + \Delta, \label{eq_delay}
\end{align}
where $[\cdot]_i$ denotes the $i$-th entry of a vector, $c$ is the speed of light, and $\Delta$ is the unknown clock bias between the \ac{5g} \ac{bs} and the user.
We further stack these \ac{5g} observations as 
$
    \thetav^{\text{az}} = [\theta^{\text{az}}_1,\theta^{\text{az}}_2,\dots,\theta^{\text{az}}_L]^\TT,\
    \thetav^{\text{el}} = [\theta^{\text{el}}_1,\theta^{\text{el}}_2,\dots,\theta^{\text{el}}_L]^\TT,\
    \tauv = [\tau_1,\tau_2,\dots,\tau_L]^\TT,
$
and finally in one vector as 
\begin{equation}\label{eq_5Gobs}
    \yv_2 = [(\thetav^{\text{az}})^\TT,(\thetav^{\text{el}})^\TT,\tauv^\TT]^\TT\in\mathbb{R}^{3L}.
\end{equation}

\subsection{Problem Formulation}
For a joint \ac{5g}-\ac{rtk} localization formulation, suppose we have two noisy observation vectors $\hat{\yv}_1$ and $\hat{\yv}_2$,
and the corresponding covariance matrices are available and denoted as $\Qm_{\hat{\yv}_1}$ and $\Qm_{\hat{\yv}_2}$.
The unknown parameters are defined as 
\begin{equation}\label{eq_x}
    \xv \triangleq [\pv_u^\TT,\kv^\TT,\Delta]^\TT\in\mathbb{R}^{N+3}.
\end{equation} 
Based on the developed \ac{rtk} model~\eqref{eq_RTKobs} and \ac{5g} model~\eqref{eq_aodaz}--\eqref{eq_delay}, we can construct the following optimization problem 
\begin{equation}\label{eq_argmin}
    \argmin_{\pv_u\in\mathbb{R}^{3}, \kv \in \mathbb{Z}^{N-1}, \Delta\in\mathbb{R}} \!  \epsilon\|\hat{\yv}_1-\Am\xv+\bv\|^2_{\Wm_1}
    + (1-\epsilon) \! \|\hat{\yv}_2-\yv_2(\xv)\|^2_{\Wm_2},
\end{equation}
where $\epsilon\in[0,1]$ is a weighting factor, and $\|(\cdot)\|_\Wm^2 \triangleq (\cdot)^\TT\Wm(\cdot)$.
The matrix $\Am$ is defined as 
\begin{align}
    \Am \triangleq \begin{bmatrix}
        \Hm & \Zerom & \zerov\\
        \Hm & \lambda\Id & \zerov\\
    \end{bmatrix}\in\mathbb{R}^{(2N-2)\times(N+3)},
\end{align}
and the nonlinear function $\yv_2(\xv)$ are represented by~\eqref{eq_aodaz}--\eqref{eq_delay}.

In this work, we use the weight matrices $\Wm_1$ and $\Wm_2$ as
\begin{align}
    \Wm_1 = \Qm_{\hat{\yv}_1}^{-1}/\|\Qm_{\hat{\yv}_1}^{-1}\|_\mathrm{F},\quad
    \Wm_2 = \Qm_{\hat{\yv}_2}^{-1}/\|\Qm_{\hat{\yv}_1}^{-1}\|_\mathrm{F}, 
\end{align}
where $\|\cdot\|_\mathrm{F}$ stands for the Frobenius norm.

\section{Methodology}
This section proposes a gradient-based solution for~\eqref{eq_argmin} and localization availability is discussed.
To be clear, we start from the typical \ac{rtk} routine with integer least-squares (ILS), based on which the proposed algorithm is developed.

\subsection{ILS-based RTK Solution}\label{sec_RTKsolution}

According to the standalone \ac{rtk} model~\eqref{eq_RTKobs}, we have the following (mixed) ILS problem \cite{teunissen2017springer}:
\begin{equation}\label{eq_MILS}
    \argmin_{\pv_u \in\mathbb{R}^{3}, \kv \in \mathbb{Z}^{N-1}} \|\yv_1-\Bm\pv_u -\Cm\kv +\bv\|^2_{\Wm_1}.
\end{equation}
It is not straightforward to solve \eqref{eq_MILS} due to the presence of the integer constraint on $\kv$. To simplify the problem, one can first ignore the constraint to obtain a \textit{float} solution as a starting point to perform the integer search. Based on the \ac{ls} estimation principle, the float solution reads
\begin{equation}\label{eq_FloatSol}
    \begin{bmatrix}
        \hat{\pv}_u \\
        \hat{\kv} 
    \end{bmatrix}
   \! =  \!\begin{bmatrix}
        \Bm^\TT \Wm_1 \Bm & \Bm^\TT \Wm_1 \Cm\\
        \Cm^\TT \Wm_1 \Bm & \Cm^\TT \Wm_1 \Cm
    \end{bmatrix}^{-1}
    \begin{bmatrix}
        \Bm^\TT \Wm_1 \left( \yv_1  \!+ \! \bv\right)\\
        \Cm^\TT \Wm_1 \left( \yv_1  \!+ \! \bv\right)
    \end{bmatrix},
\end{equation}
with the covariance matrix given by
\begin{equation}\label{eq_FloatVar}
    \begin{bmatrix}
        \Qm_{\hat{\pv}_u}\! & \!\Qm_{\hat{\pv}_u\hat{\kv}} \\
        \Qm_{\hat{\kv}\hat{\pv}_u}\! &\! \Qm_{\hat{\kv}}
    \end{bmatrix}
   \! =  \!\begin{bmatrix}
        \Bm^\TT \Wm_1 \Bm & \Bm^\TT \Wm_1 \Cm\\
        \Cm^\TT \Wm_1 \Bm & \Cm^\TT \Wm_1 \Cm\\
    \end{bmatrix}^{-1}.
\end{equation}

The objective function in \eqref{eq_MILS} can be decomposed into three easy-to-evaluate terms as \cite{teunnissen1995least}
\begin{equation} \label{eq_Decompose}
\resizebox{.89\hsize}{!}{
$
\begin{aligned}
    &\|\yv_1  \! -  \! \Bm\pv_u  \! -  \! \Cm\kv  \! + \! \bv\|^2_{\Wm_1} \\ = &\|\yv_1  \! -  \! \Bm\hat{\pv}_u  \! -  \! \Cm \hat{\kv}  \! +  \! \bv\|^2_{\Wm_1}  \! + \!  \|\hat{\kv}  \! -  \! \kv\|^2_{\Qm_{\hat{\kv}}^{-1}}  \! +  \! \|\hat{\pv}_u  \! \left(\kv \right)  \! -  \! \pv_u\|^2_{\Qm_{\hat{\pv}_u \! \left(\kv \right)}^{-1}},
\end{aligned}
$}
\end{equation}
where $\hat{\pv}_u  \! \left(\kv \right)$ represents the \ac{ls} solution conditioned on $\kv$. That is
\begin{equation} \label{eq_LSCondition}
    \hat{\pv}_u  \! \left(\kv \right) = \hat{\pv}_u - \Qm_{\hat{\kv}\hat{\pv}_u}\Qm_{\hat{\kv}}^{-1}\left(\hat{\kv}-\kv\right),
\end{equation}
and the corresponding covariance matrix is 
$
    \Qm_{\hat{\pv}_u \! \left(\kv \right)} = \Qm_{\hat{\pv}_u} - \Qm_{\hat{\pv}_u\hat{\kv}}\Qm_{\hat{\kv}}^{-1}\Qm_{\hat{\kv}\hat{\pv}_u}.
$
The first term on the right-hand side of \eqref{eq_Decompose} is given in a close-form. The last term of \eqref{eq_Decompose} is irrelevant to the integer search and can be eliminated \cite{teunnissen1995least}. Therefore, the optimization \eqref{eq_MILS} reduces to
\begin{equation} \label{eq_ILS}
    \check{\kv} = \arg \min_{\kv \in \mathbb{Z}^{N-1}} \|\hat{\kv}  \! -  \! \kv\|^2_{\Qm_{\hat{\kv}}^{-1}},
\end{equation}
The LAMBDA method is usually utilized to resolve the unknown integers in \eqref{eq_ILS} because of its high computational efficiency and capacity to maximize the success rate \cite{teunnissen1995least}.  
Once the integer ambiguities are resolved, the updated receiver position is given by $ \check{\pv}_u = \hat{\pv}_u  \left(\check{\kv} \right)$. 

\subsection{The Proposed Hybrid GNSS-5G Localization Algorithm}
\label{sec_jointSol}
\subsubsection{Initialization}
Since~\eqref{eq_argmin} is a non-convex optimization problem, proper initialization is essential to avoid local minima.
In general, one can consider using the float RTK solution in~\eqref{eq_FloatSol} as an initial user position and float carrier-phase ambiguities, and the  clock bias can be randomly initialized from a uniform distribution as $\tilde{\Delta}_0\sim\mathcal{U}(0,T_c)$ with $T_c$ is the clock cycle of the user, that is 
\begin{equation}\label{eq_init4}
    \xv_0 = [\hat{\pv}_{u}^\TT,\hat{\kv}^\TT,\tilde{\Delta}_0]^\TT.
\end{equation}
However, the standalone \ac{rtk} solution may not be available in some cases, as will be discussed in the next subsection.
Under these circumstances, we can initialize based on the \ac{5g} observations using
\begin{equation}\label{eq_init3}
    \xv_0 = [\hat{\pv}_{u,0}^\TT,\hat{\kv}_0^\TT, \tilde{\Delta}_0]^\TT.
\end{equation}
Here, $\hat{\pv}_{u,0}$ is estimated as
\begin{equation}\label{eq_pu0}
    \hat{\pv}_{u,0} = \frac{1}{L}\sum_{\ell=1}^L \left( \pv_{\Bt,\ell} + c\tau_\ell\Rm_{\Bt,\ell}\tv_{\ell} \right),
\end{equation}
where 
$
    \tv_{\ell} =
  \begin{bmatrix}
    \cos(\theta_\ell^\text{az})\cos(\theta_\ell^\text{el}),
    \sin(\theta_\ell^\text{az})\cos(\theta_\ell^\text{el}), 
    \sin(\theta_\ell^\text{el})
  \end{bmatrix}^\TT.
$
The float carrier-phase ambiguities $\hat{\kv}_0$ can be estimated based on the \ac{ls} solution of~\eqref{eq_MILS} given $\pv_u=\hat{\pv}_{u,0}$, i.e., 
\begin{equation}\label{eq_Initial}
        \hat{\kv}_0 
   \! =  \left(\Cm^\TT \Wm_1 \Cm \right)^{-1}
        \Cm^\TT \Wm_1 \left( \yv_1  \!- \!\Bm\hat{\pv}_{u,0}\!+ \! \bv\right).
\end{equation}

\subsubsection{Float Solution}
With proper initialization, we can obtain a float solution of~\eqref{eq_argmin} by ignoring the integer constraint on $\kv$.
In this case the problem~\eqref{eq_argmin} is reduced to 
\begin{equation}\label{eq_argmin2}
    \argmin_{\xv\in\mathbb{R}^{N+3}}\ \epsilon\|\hat{\yv}_1-\Am\xv+\bv\|^2_{\Wm_1}
    + (1-\epsilon)\|\hat{\yv}_2-\yv_2(\xv)\|^2_{\Wm_2}.
\end{equation}
A gradient-based algorithm (such as gradient descent) can be applied to solve~\eqref{eq_argmin2}.
We define $f_1 \triangleq \|\hat{\yv}_1-\Am\xv+\bv\|^2_{\Wm_1}$ and $f_2 \triangleq \|\hat{\yv}_2-\yv_2(\xv)\|^2_{\Wm_2}$.
The first-order derivatives $\partial f_1/\partial \xv$ and $\partial f_2/\partial \xv$ for the iterative algorithm are given by
\begin{align}
    {\partial f_1}/{\partial \xv} &= -2\Am^\TT\Wm_1(\hat{\yv}_1-\Am\xv+\bv),\\
    {\partial f_2}/{\partial \xv} &= -2\left(\frac{\partial \yv_2(\xv)}{\partial \xv}\right)^\TT\Wm_2\left(\hat{\yv}_2 - \yv_2(\xv)\right).
\end{align}
The expressions of ${\partial \yv_2(\xv)}/{\partial \xv}$ can be collected from the following derivatives:
\begin{align}
    \notag
    &{\partial \theta_\ell^\text{az}}/{\partial \pv_u} = s_1 \frac{(\uv_1^\TT\tv_\ell)\Rm_{\Bt,\ell}\uv_2 - (\uv_2^\TT\tv_\ell)\Rm_{\Bt,\ell}\uv_1}{(\uv_1^\TT\tv_\ell)^2},\\
    \notag
    &{\partial \theta_\ell^\text{el}}/{\partial \pv_u} = s_2 \left(\frac{\Rm_{\Bt,\ell}\uv_3}{d_\ell} - \frac{(\uv_3^\TT\tv_\ell)(\pv_u-\pv_{\Bt,\ell})}{d_\ell^3}\right),\\
    \notag
    &{\partial \tau_\ell}/{\partial \pv_u} = \frac{\pv_u - \pv_{\Bt,\ell}}{cd_\ell},\quad \frac{\partial \tau_\ell}{\partial \Delta} = 1,
\end{align}
where
\begin{align}
    \notag
    &\uv_1 = [1,0,0]^\TT,\  \uv_2 = [0,1,0]^\TT,\  \uv_3 = [0,0,1]^\TT, \\
    \notag
    &d_\ell = \|\pv_u - \pv_{\Bt,\ell}\|_2,\quad \tv_\ell = \Rm_{\Bt,\ell}^\TT(\pv_u - \pv_{\Bt,\ell}),\\
    \notag
    &s_1 = \left[1 + \left(\frac{ \mathbf{u}_2^\mathsf{T} \tv_\ell }{ \mathbf{u}_1^\mathsf{T} \tv_\ell }\right)^2\right]^{-1},\ 
    s_2 = \left[1 - \left(\frac{ \mathbf{u}_3^\mathsf{T} \tv_\ell }{ d_\ell }\right)^2\right]^{-\frac{1}{2}}.\notag
\end{align}

\subsubsection{Ambiguity Resolution}
Suppose a float solution $\hat{\xv} = [\hat{\pv}_{u}^\TT,\hat{\kv}^\TT,\hat{\Delta}]^\TT$ is obtained by solving~\eqref{eq_argmin2}.
The integer carrier-phase ambiguity resolution step can be achieved by the same routine as Subsection~\ref{sec_RTKsolution}, i.e., solving~\eqref{eq_ILS} using, e.g., the LAMBDA methods.
Therefore, the integer ambiguity estimate is obtained as $\check{\kv}$.

\subsubsection{Fixed Solution}
After the ambiguity resolution, the fixed solution can be obtained through the following optimization:
\begin{equation}\label{eq_argmin3}
    \argmin_{\pv_u\in\mathbb{R}^{3}, \Delta\in\mathbb{R}}\ \epsilon\|\hat{\yv}_1-\Am\xv+\bv\|^2_{\Wm_1}
    + (1-\epsilon)\|\hat{\yv}_2-\yv_2(\xv)\|^2_{\Wm_2},
\end{equation}
with a fixed $\kv = \check{\kv}$. The solution is returned as $\check{\pv}_u$ and $\check{\Delta}$.
Finally, we have the joint estimate as $\check{\xv} = [\check{\pv}_u^\TT, \check{\kv}^\TT, \check{\Delta}]^\TT$.

\subsection{Localization Availability Analysis}
Generally, the \ac{rtk} technique requires at least four satellites to perform localization.
However, by incorporating the \ac{5g} observations, this constraint can be further relaxed.
The localization availability can be determined by comparing the dimensionality of the observations and that of the unknowns.
In general, the number of observations should be equal to or greater than the number of unknowns to make the estimation problem can be solved with a unique solution.
In this paper, we name the cases with the localization uniqueness (i.e., the observations' dimension is not less than unknowns) as the \textit{localizable} cases, otherwise are \textit{nonlocalizable} cases.
Taking the case where $N=3$ and $L=1$ as an example, we can check the unknown entries of $\xv\in\mathbb{R}^{6}$ in~\eqref{eq_x} and the observations $[\yv_1^\TT,\yv_2^\TT]^\TT\in\mathbb{R}^{7}$ in~\eqref{eq_RTKobs} and~\eqref{eq_5Gobs}.
As the dimension of the observations is higher than the unknowns, the user is localizable when $N=3$ and $L=1$. 
A summary of the localization availability in different scenarios is presented in Table~\ref{tab1}.
We can see that leveraging \ac{5g} observations enhances the localization availability.
For example, with a single \ac{5g} \ac{bs}, localization is available in the cases of $N=2$ and $N=3$.
Note that having \ac{5g} observations from two or more \acp{bs} is sufficient to perform localization regardless of GNSS availability.

\begin{table}[t]
\renewcommand{\arraystretch}{1.5}
    \begin{center}
    \begin{threeparttable}
    \caption{Demonstration of the localization availability.\tnote{*}}
    \label{tab1}
    \setlength{\tabcolsep}{1.8mm}{
        \begin{tabular}{|c|c|c|c|c|c|c|}
        \hline
        \diagbox{$L$}{($n_o$, $n_u$)}{$N$} & 0 & 1 & 2 & 3 & 4 & 5\\
        \hline\hline
        0 & \sout{$(0,3)$} & \sout{$(0,3)$} & \sout{$(2,4)$} & \sout{$(4,5)$} & $(6,6)$ & $(8,7)$  \\
        \hline
        1 & \sout{$(3,4)$} & \sout{$(3,4)$} & $(5,5)$ & $(7,6)$ & $(9,7)$ & $(11,8)$  \\
        \hline
        \end{tabular}
    }
        \begin{tablenotes}
            \footnotesize
            \item[*] Here $n_o$ and $n_u$ represent the number of observations and unknowns, respectively. The nonlocalizable cases ($n_o<n_u$) are marked by a strikethrough.
        \end{tablenotes}
    \end{threeparttable}
    \end{center}
\end{table}

\section{Performance Evaluation}

\subsection{Simulation Setup}

The simulations are implemented using the actual satellite orbit information in the GPS Yuma Almanacs file on January 01, 2023.
The positions and the orientations of the \ac{5g} \acp{bs} are generated randomly within a $\unit[50]{m}\times\unit[50]{m}\times\unit[50]{m}$ space.
We assume the observations $\yv_1=[\pv^\TT,\phiv^\TT]^\TT$ and $\yv_2$ to be contaminated with zero-mean additive Gaussian noises controlled by their standard deviations. We set the standard deviation of the carrier-phase measurements equal to a value $\sigma$ and that of the pseudo-range data equal to $100\sigma$. The standard deviation of the noise of $\yv_2$ is fixed based on the Fisher information matrix of the \ac{5g} channel estimation step, as detailed in, e.g.,~\cite{Zheng2022Misspecified,Chen2022Tutorial,Zheng2022Coverage}.
In the cases where $L=0$, the results are obtained from the ILS solution in Subsection~\ref{sec_RTKsolution}. In all other cases, the proposed method developed in Subsection~\ref{sec_jointSol} is used with a weighting factor $\epsilon=0.6$. The iterative procedure for solving~\eqref{eq_argmin} is implemented using the Manopt toolbox~\cite{manopt}.
All the involved \acp{rmse} are computed through 500 Monte Carlo simulations.

\subsection{Results Analysis}

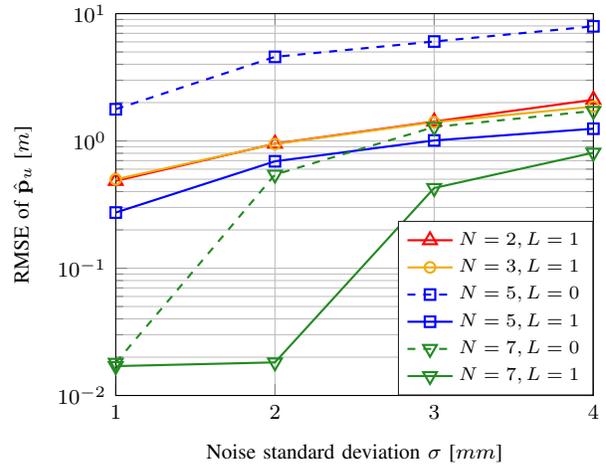
\begin{figure}[t]
    \begin{minipage}[b]{0.1\linewidth}
      \centering
\definecolor{ForestGreen}{rgb}{0.1333    0.5451    0.1333}%
\definecolor{DarkGoldenrod}{rgb}{0.9333    0.6784    0.0549}%
\definecolor{BlueViolet}{rgb}{ 0.6275    0.1255    0.9412}%
\definecolor{Firebrick}{rgb}{0.8039    0.1490    0.1490}
\begin{tikzpicture}

\begin{axis}[%
width=2.5in,
height=2in,
at={(0in,0in)},
scale only axis,
xmin=1,
xmax=4,
xtick={1,2,3,4},
xticklabel style = {font=\color{white!15!black},font=\footnotesize},
xlabel style={font=\color{white!15!black},font=\footnotesize},
xlabel={Noise standard deviation $\sigma$ [$mm$]},
ymode=log,
ymin=0.01,
ymax=10,
yminorticks=true,
yticklabel style = {font=\color{white!15!black},font=\footnotesize},
ylabel style={font=\color{white!15!black},font=\footnotesize},
ylabel={RMSE of $\hat{\pv}_u$ [$m$]},
axis background/.style={fill=white},
xmajorgrids,
xminorgrids,
ymajorgrids,
yminorgrids,
legend style={at={(1,0)}, anchor=south east, legend cell align=left, align=left, draw=white!15!black, font=\scriptsize}
]
\addplot [color=red, mark=triangle, line width=0.8pt, mark options={solid, red}, mark size=3pt]
  table[row sep=crcr]{%
1	0.485202265430153\\
2	0.950962688036195\\
3	1.42010666783225\\
4	2.10653517742219\\
};
\addlegendentry{$N=2,L=1$}

\addplot [color=DarkGoldenrod, mark=o, line width=0.8pt, mark options={solid, DarkGoldenrod}]
  table[row sep=crcr]{%
1	0.498820696806783\\
2	0.945652454901207\\
3	1.40499259082266\\
4	1.86117952214781\\
};
\addlegendentry{$N=3,L=1$}

\addplot [color=blue, dashed, mark=square, line width=0.8pt, mark options={solid, blue}]
  table[row sep=crcr]{%
1	1.77443134740579\\
2	4.57580201891068\\
3	6.04071398184584\\
4	7.94788812905881\\
};
\addlegendentry{$N=5,L=0$}

\addplot [color=blue, mark=square, line width=0.8pt, mark options={solid, blue}]
  table[row sep=crcr]{%
1	0.273642358746277\\
2	0.691697275196753\\
3	1.0088368881088\\
4	1.2469108524401\\
};
\addlegendentry{$N=5,L=1$}

\addplot [color=ForestGreen, dashed, mark=triangle, line width=0.8pt, mark options={solid, rotate=180, ForestGreen}, mark size=3pt]
  table[row sep=crcr]{%
1	0.0180684608215257\\
2	0.543474679350813\\
3	1.28709122320313\\
4	1.72061513612301\\
};
\addlegendentry{$N=7,L=0$}

\addplot [color=ForestGreen, mark=triangle, line width=0.8pt, mark options={solid, rotate=180, ForestGreen}, mark size=3pt]
  table[row sep=crcr]{%
1	0.0170070474356009\\
2	0.0181979827700056\\
3	0.426597891470954\\
4	0.808580451050949\\
};
\addlegendentry{$N=7,L=1$}

\end{axis}
\end{tikzpicture}%
        \vspace{-2em}
    \end{minipage}
    \caption{
        Evaluation of positioning \ac{rmse} versus the noise standard deviation of the carrier-phase noise for different  numbers of satellites $N$ and $L=\{1,0\}$.
    }
    \label{fig_1}
    \vspace{-3mm}
\end{figure}

Fig.~\ref{fig_1} presents the \ac{rmse} of estimated $\hat{\pv}_u$ versus the carrier-phase noise standard deviation $\sigma$ for different  number of satellites $N$ and fixed $L=\{1,0\}$.
In general, we can see that the \ac{rmse} increases as the noise level increases.
By comparing the cases of $L=0$ (dashed curves) and $L=1$ (solid curves) for the same $N$, we observe that adding one \ac{5g} \ac{bs} offers a significant reduction in estimation error, 
demonstrating that the utilization of the \ac{5g} observations can provide a remarkable improvement in localization performance. 
Moreover, it is noted that with the \ac{5g} observations involved, localization in the cases where $N=2$ and $N=3$ become not only localizable but also with higher accuracy than the case with $5$-satellite and no 5G aid (dashed blue curve).
In addition, by comparing the cases of different $N$ with the same $L$, we observe that the more satellites available, the lower the estimation error is. It is also observed that the performance of the cases $N=2$ and $N=3$ is very close, indicating that under a circumstance with insufficient satellites ($N<4$), the location information is mainly derived from the \ac{5g} observations and changing the number of satellite within $N<4$ cannot boost performance significantly.

\begin{figure}[t]
    \begin{minipage}[b]{0.1\linewidth}
      \centering
\definecolor{ForestGreen}{rgb}{0.1333    0.5451    0.1333}%
\definecolor{DarkGoldenrod}{rgb}{0.9333    0.6784    0.0549}%
\definecolor{BlueViolet}{rgb}{ 0.6275    0.1255    0.9412}%
\definecolor{Firebrick}{rgb}{0.8039    0.1490    0.1490}
\begin{tikzpicture}

\begin{axis}[%
  width=2.5in,
  height=1.7in,
  at={(0in,0in)},
scale only axis,
xmin=1,
xmax=4,
xtick={1,2,3,4},
xticklabel style = {font=\color{white!15!black},font=\footnotesize},
xlabel style={font=\color{white!15!black},font=\footnotesize},
xlabel={Noise standard deviation $\sigma$ [$mm$]},
ymode=log,
ymin=0.01,
ymax=10,
yticklabel style = {font=\color{white!15!black},font=\footnotesize},
ylabel style={font=\color{white!15!black},font=\footnotesize},
ylabel={RMSE of $\hat{\pv}_u$ [$m$]},
yminorticks=true,
axis background/.style={fill=white},
xmajorgrids,
xminorgrids,
ymajorgrids,
yminorgrids,
legend style={at={(0.995,1.01)}, anchor=south east, legend cell align=left, legend columns=2, align=left, draw=white!15!black, font=\scriptsize}
]
\addplot [color=ForestGreen, dashed, mark=triangle, line width=0.8pt, mark options={solid, ForestGreen}, mark size=3pt]
  table[row sep=crcr]{%
1	1.7803045714078\\
2	4.49672040535962\\
3	6.08969403733626\\
4	8.6220212171885\\
};
\addlegendentry{$\ N=5\ ,\  L=0$\qquad\quad\ }

\addplot [color=ForestGreen, mark=triangle, line width=0.8pt, mark options={solid, ForestGreen}, mark size=3pt]
  table[row sep=crcr]{%
1	0.0180668096192474\\
2	0.524731106921216\\
3	1.27065820055572\\
4	1.86649244693156\\
};
\addlegendentry{$\ N=7\ ,\  L=0$}

\addplot [color=blue, dashed, line width=0.8pt, mark=square, mark options={solid, blue}]
  table[row sep=crcr]{%
1	0.253502124841791\\
2	0.719430319601798\\
3	1.00500502756721\\
4	1.2543887957282\\
};
\addlegendentry{$\ N=5\ ,\  L=1$}

\addplot [color=blue, mark=square, line width=0.8pt, mark options={solid, blue}]
  table[row sep=crcr]{%
1	0.01698820954486\\
2	0.0569994783006026\\
3	0.351850690008989\\
4	0.798620555746058\\
};
\addlegendentry{$\ N=7\ ,\  L=1$}

\addplot [color=DarkGoldenrod, dashed, mark=o, line width=0.8pt, mark options={solid, DarkGoldenrod}]
  table[row sep=crcr]{%
1	0.212347170987487\\
2	0.345749582635028\\
3	0.423776362262728\\
4	0.480014574426533\\
};
\addlegendentry{$\ N=5\ ,\  L=2$}

\addplot [color=DarkGoldenrod, mark=o, line width=0.8pt, mark options={solid, DarkGoldenrod}]
  table[row sep=crcr]{%
1	0.01651939755578\\
2	0.0176681729674025\\
3	0.0951648313585418\\
4	0.255107898588521\\
};
\addlegendentry{$\ N=7\ ,\  L=2$}

\addplot [color=red, dashed, mark=triangle, line width=0.8pt, mark options={solid, rotate=180, red}, mark size=3pt]
  table[row sep=crcr]{%
1	0.209149980978298\\
2	0.329969270429942\\
3	0.399307091413625\\
4	0.456416228240311\\
};
\addlegendentry{$\ N=5\ ,\  L=3$}

\addplot [color=red, mark=triangle, line width=0.8pt, mark options={solid, rotate=180, red}, mark size=3pt]
  table[row sep=crcr]{%
1	0.0165303721919321\\
2	0.0176511398183167\\
3	0.083753574137448\\
4	0.229134113220739\\
};
\addlegendentry{$\ N=7\ ,\  L=3$}

\end{axis}
\end{tikzpicture}%
        \vspace{-2em}
    \end{minipage}
    \caption{Positioning \ac{rmse} versus carrier-phase noise standard deviation for different numbers of \ac{5g} \ac{bs} $L\in\{0,1,2,3\}$ and different numbers of satellites $N\in\{5,7\}$).}
    \label{fig_2}
    \vspace{-3mm}
\end{figure}
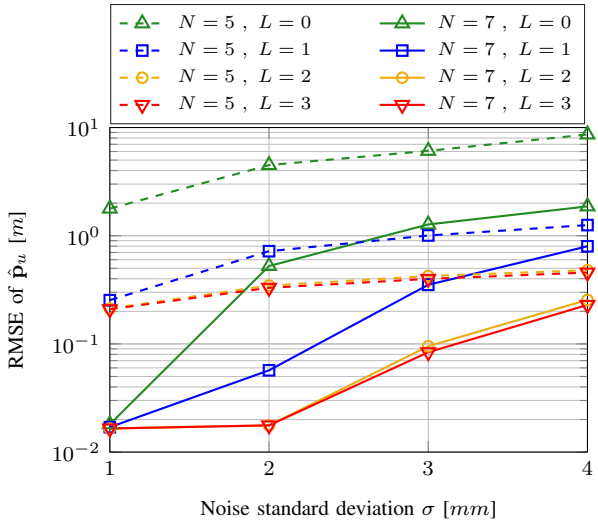

Fig.~\ref{fig_2} plots the \ac{rmse} of $\hat{\pv}_u$ versus the carrier-phase noise standard deviation $\sigma$ for different numbers of \ac{5g} \acp{bs} $L$ and fixed $N=\{5,7\}$.
The results show that for the same number of satellites, the more \ac{5g} \acp{bs} we deploy, the better localization performance we obtain, especially for $L\in\{0,1,2\}$.
However, when more than two \ac{5g} \acp{bs} are available, continuing to increase the number of \ac{5g} \acp{bs} may not significantly improve performance.

\begin{table}[t]
    \renewcommand{\arraystretch}{1.4}
        \begin{center}
        \begin{threeparttable}
        \caption{Success Rate of the Carrier-Phase Ambiguity Resolution}
        \label{tab2}
        \setlength{\tabcolsep}{1.3mm}{
            \begin{tabular}{|c|c|c|c|c|c|c|}
            \hline
            \diagbox{$L$}{$N$} & 2 & 3 & 4 & 5 & 6 & 7 \\
            \hline\hline
            0 & $\unit[0.00]{\%}$ & $\unit[0.00]{\%}$ & $\unit[7.02]{\%}$ & $\unit[54.39]{\%}$ & $\unit[94.66]{\%}$ & $\unit[99.99]{\%}$  \\
            \hline
            1 & $\unit[36.08]{\%}$ & $\unit[45.17]{\%}$ & $\unit[51.24]{\%}$ & $\unit[73.58]{\%}$ & $\unit[99.06]{\%}$ & $\unit[100.00]{\%}$  \\
            \hline
            \end{tabular}
        }
        \end{threeparttable}
        \end{center}
 \end{table}

Finally, we evaluate the success rate of carrier-phase ambiguity resolution, an important performance indicator for GNSS-based positioning.
Performance is computed from 10000 Monte Carlo trials for each case. The results are shown in Table~\ref{tab2}.
It is clearly visible that the introduction of \ac{5g} observations can increase the ambiguity resolution success rate.
Additionally, in general, access to more satellite observations is also helpful.

\section{Conclusion} 

This paper formulated, analyzed, and solved a \ac{5g}-aided \ac{gnss} localization problem in satellite-deprived environments.
A novel gradient-based algorithm, coupled with a proposed ambiguity resolution method, is applied to estimate the user's location using \ac{gnss} and \ac{5g} observations simultaneously.
The presented results reveal that the proposed approach enhances localization accuracy and GNSS ambiguity resolution success rates, especially in scenarios with extremely limited satellite visibility.

\normalem
\bibliographystyle{IEEEtran}
\bibliography{references}

\end{document}